\documentclass[a4paper,12pt,nohyper]{JHEP3}
\usepackage{epsfig}
\usepackage{amsmath,latexsym,amssymb}
\usepackage{graphicx}
\usepackage[latin1]{inputenc}
\usepackage{braket}

\usepackage{framed}

\def\be{\begin{equation}}
\def\ee{\end{equation}}
\def\ba{\begin{eqnarray}}
\def\ea{\end{eqnarray}}

\newcommand{\bn}[2]{\left(\begin{array}{c} #1\\ #2
\end{array}\right)}

\newcommand{\da}{\dot{\alpha}}
\newcommand{\dbe}{\dot{\beta}}

\def\bra#1{\mathinner{\langle{#1}|}}
\def\ket#1{\mathinner{|{#1}\rangle}}
\def\sbra#1{\mathinner{\lbrack{#1}|}}
\def\sket#1{\mathinner{|{#1}\rbrack}}
\def\braket#1{\mathinner{\langle{#1}\rangle}}
\newcommand{\sbraket}[1]{\lbrack #1\rbrack}

\newcommand{\la}{\lambda}
\newcommand{\tla}{\tilde{\lambda}}
\newcommand{\ts}{\tilde{s}}
\newcommand{\tit}{\tilde{t}}
\newcommand{\tu}{\tilde{u}}

\newcommand{\apr}{\alpha'}
\newcommand{\hapr}{{\apr \over 2}}

\newcommand{\hk}{\hat{k}}
\newcommand{\has}{\hat{s}}
\newcommand{\he}{\hat{e}}

\def\a{\alpha}

\def\b{\beta}

\def\e{\epsilon}

\def\m{\mu}

\title{\boldmath BCFW construction of the Veneziano Amplitude\unboldmath}

\author{Angelos~Fotopoulos${}^{\spadesuit}$ \\

  \begin{itemize}
    
  \item  Dipartimento di Fisica Teorica
    dell Universita di Torino and\\
  INFN Sezione di Torino,\\
  via P. Giuria 1, I-10125 Torino, Italy.
  
  \end{itemize}

\bigskip

E-mail: \email{foto@to.infn.it}}

\abstract{In this note we demonstrate how one can compute the
Veneziano amplitude for bosonic string theory using the BCFW
method. We use an educated ansatz for the cubic amplitude of two
tachyons and an arbitrary level string state.}

\preprint{DFTT-22/2010}

\begin{document}

\setcounter{figure}{0} \setcounter{table}{0}

\setcounter{footnote}{0}
\renewcommand{\thefootnote}{\arabic{footnote}}
\setcounter{section}{0}

\section{Introduction} \label{intro}
Recently, there has been remarkable progress in exploring the
properties of the S-matrix for tree level scattering amplitudes in
gauge and gravity theories. Motivated by Witten's twistor
formulation of ${\cal N}=4$ Super--Yang--Mills (SYM)
\cite{Witten:2003nn}, several new methods have emerged which allow
one to compute tree level amplitudes. The Cachazo--Svrcek--Witten
(CSW)  method  \cite{Cachazo:2004kj} has demonstrated how one can
use the maximum helicity violating (MHV) amplitudes of
\cite{Parke:1986gb} as field theory vertices to construct
arbitrary gluon amplitudes.

Analyticity of gauge theory tree level amplitudes has lead to the
Britto--Cachazo--Feng--Witten (BCFW) recursion relations
\cite{Britto:2004ap, Britto:2005fq, Cachazo:2005ga}. Specifically,
analytic continuation of external momenta in a scattering
amplitude allows one, under certain assumptions, to determine the
amplitude through its residues on the complex plane. Locality and
unitarity require that the residue at the poles is a product of
lower-point amplitudes. Therefore one obtains recursion relation
of higher point amplitudes as a product of lower point ones alas
computed at complex on-shell momenta. Actually the CSW
construction turns out to be a particular application of the BCFW
method \cite{Risager:2005vk}.

The power of these new methods extends beyond computing
 tree level amplitudes. The
original recursion relations for gluons \cite{Britto:2004ap} were
inspired by the infrared (IR) singular behavior of ${\cal N} = 4$
SYM. Tree amplitudes for the emission of a soft gluon from a given
n-particle process are IR divergent and this divergence is
cancelled by IR divergences from soft gluons in the 1-loop
correction. For maximally supersymmetric theories these IR
divergencies suffice to determine fully the form of the 1-loop
amplitude. Therefore, there is a direct link between tree level
and loop amplitudes. Recently there has been intense investigation
towards a conjecture \cite{ArkaniHamed:2009dn, ArkaniHamed:2009si}
which relates IR divergencies of multiloop amplitudes with those
of lower loops, allowing therefore the analysis of the full
perturbative expansion of these gauge theories\footnote{Very
recently there has been a proposal for the amplitude integrand at
any loop order in ${\cal N}=4$ SYM \cite{ArkaniHamed:2010kv} and a
similar discussion on the behavior of loop amplitudes under BCFW
deformations \cite{Boels:2010nw}.}. These new methods have
revealed a deep structure hidden in maximally supersymmetric gauge
theories \cite{ArkaniHamed:2008gz} and possibly in more general
gauge theories and gravity.

The recursion relations of \cite{Britto:2005fq} are in the heart
of many of the aforementioned developments. Nevertheless it
crucially relies on the asymptotic behavior of amplitudes under
complex deformation of some external momenta. When the complex
parameter, which parametrizes the deformation, is taken to
infinity an amplitude should fall sufficiently fast so that there
is no pole at infinity\footnote{There has been though some recent
progress  \cite{Feng:2009ei,Feng:2010ku}
 in generalizing the  BCFW relations for theories with boundary
contributions.}. Although naive power counting of individual
Feynman diagrams seems to lead to badly divergent amplitudes for
large complex momenta, it is intricate cancellations among them
which result in a much softer behavior than expected. Gauge
invariance and supersymmetry in some cases lies into the heart of
these cancellations.

It is natural to wonder whether these field theoretic methods can
be applied and shed some light into the structure of string theory
amplitudes. This is motivated, in particular, by the fact that the
theory that plays
 a central role in the developments we described above, that is
${\cal N}=4$ SYM,  appears as the low-energy limit of  string
theory in the presence of  D3-branes. Moreover there is a constant
interest on string amplitudes over the years \cite{Garousi:1996ad}
in principle for lower point amplitudes used to derive features of
the low energy effective actions in string theory. Recently there
is an ongoing effort to derive formulas for arbitrary n-point
functions \cite{Stieberger:2006bh}, in particular the n-gluon
amplitude, and production of massive string states
\cite{Anchordoqui:2008hi} which would be useful for predictions
based on string theory in experiments like the LHC.

In order to even consider applying the aforementioned methods to
string scattering amplitudes, one needs as a first step to study
their behavior for large complex momenta. Since generally string
amplitudes are known to have excellent large momentum behavior,
one expects that recursion relations should be applicable here as
well. Nevertheless, one should keep in mind that although the
asymptotic amplitude behavior might be better than any local field
theory, the actual recursion relations will be quite more
involved. The reason is that they will require knowledge of an
infinite set of on-shell string amplitudes, at least the three
point functions, between arbitrary Regge trajectory states of
string theory.

The study of the asymptotic behavior of string amplitudes under
complex momentum deformations was initiated in \cite{Boels:2008fc}
and  elaborated further in \cite{Cheung:2010vn,Boels:2010bv}.
These works established, using direct study of the amplitudes in
parallel with pomeron technics \cite{Brower:2006ea}, that both
open and closed bosonic and supersymmetric string theories have
good behavior asymptotically, therefore allowing one to use the
BCFW method. In \cite{Fotopoulos:2010cm} it was shown that the
same conclusion holds for string amplitudes in the presence of
D-branes. Recently, pomerons have also been used in
\cite{Fotopoulos:2010ay} to advocate that BCFW relations exist in
higher spin theories constructed as the tensionless limit of
string theories. Therefore it seems there is a plethora of
theories which allow under mild assumptions BCFW recursion
relations.

Moreover, it was observed in \cite{Cheung:2010vn} that for the
supersymmetric theories the leading and subleading asymptotic
behavior of open and closed string amplitudes is the same as the
asymptotic behavior of their field theory limits, i.e.~gauge and
gravity theories respectively. This led to the eikonal Regge (ER)
regime conjecture \cite{Cheung:2010vn} which
 states that string theory amplitudes, in a region where some
of the kinematic variables are much greater than the string scale
and the rest much smaller, are reproduced by their corresponding
field theory limits. For bosonic theories there is some
discrepancy in some subleading terms \cite{Boels:2010bv} which
most probably can be attributed to the fact that an effective
field theory  for bosonic strings is plagued by ambiguities due to
the presence of tachyonic modes. In \cite{Fotopoulos:2010cm} it
was shown that this conjecture does not seem to hold for mixed
open-closed string amplitudes. Nevertheless for pure open string
amplitudes this conjecture was strengthened in
\cite{Garousi:2010er} by direct calculation of n-point amplitudes
for some restricted kinematic setups. It seems that string theory
amplitudes are intimately connected to field theory amplitudes
even away from their field theory limit $\apr \to 0$.

The purpose of this work is to actually give a first example of
BCFW recursion relations for bosonic string theory. In
\cite{Cheung:2010vn} a set of recursion relations was given for
the five tachyon scattering in bosonic string. Nevertheless their
construction is based on tachyon sub-amplitudes only. This quite
different to the spirit of BCFW for field theories where
sub-amplitudes which involve arbitrary states of the theory should
be used in constructing the recursion relations. We will start
with a simpler case the four tachyon amplitude i.e. the Veneziano
amplitude \cite{Veneziano:1968yb}. We believe that this example
encompasses all the novel features of BCFW for string theory i.e.
infinite number of massive states and world-sheet duality. The
main tools we will use are application of the BCFW deformation for
massive states and an educated guess of the 3-point amplitude for
two open string tachyons and an arbitrary level string state in a
specific frame. This last ansatz is motivated by the three reggeon
vertex of \cite{Ademollo:1974kz} and recent studies of string
amplitudes like \cite{Sagnotti:2010at,Feng:2010yx}.

\section{Preliminaries}\label{Prel}
We use the massive spinor helicity formulation of
\cite{Spehler:1991yw, Boels:2010mj} in four dimensions. We analyze
any massive momentum vector $k$ in terms of two light-like momenta
$q$ and $k^\flat$
\begin{equation}\label{momdec}
k_{\mu} = k^{\flat}_{\mu} + \frac{k^2}{2 q \cdot k} q_{\mu} \ .
\end{equation}
Consistency of this equation requires
\begin{equation}
q \cdot k = q \cdot k^{\flat}
\end{equation}
A pictorial form of the formula is given in a light-cone diagram
in figure 1 \cite{Boels:2010mj}.
\begin{figure}
\begin{center} \label{momdecomp}
\includegraphics[width=5cm]{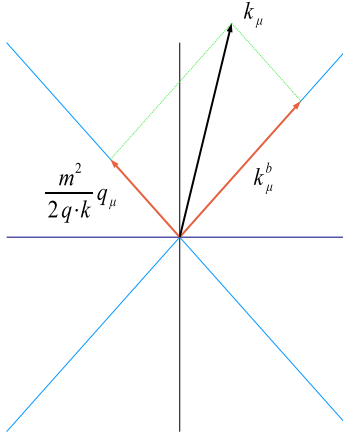}
\caption{The decomposition of an arbitrary massive vector in 4d on
two light-cone vectors \cite{Boels:2010mj}}
\end{center}
\end{figure}
If we have two massive vectors to decompose we can use the
formulas
\begin{equation}\label{2massdec}
k_1= k_1^{\flat} + {m_1^2 \over s_{12}^{\flat}} k_2^\flat\ ,
\qquad k_2= k_2^{\flat} + {m_2^2 \over s_{12}^{\flat}} k_1^\flat \
,
\end{equation}
where we define
\begin{equation}\label{mandflat}
s_{ij}^\flat = 2 k_1^\flat \cdot k_2^\flat = \braket{1 2}
\sbraket{1 2}
\end{equation}
We can define the polarization tensors for a given massive state
(gauge boson) as follows \cite{Boels:2010mj, Feng:2010yx}
\begin{equation}\label{maspolar}
e^+_{\a \da}= \sqrt{2} \ { k^\flat_{\a} q_{\da} \over \sbraket{q
k^\flat} } \ ,\quad e^-_{\a \da}= \sqrt{2} \ { q_{\a}
k^\flat_{\da} \over \braket{q k^\flat} }\ , \quad e^0_{\a \da}=
{s_{\a \da} \over m}
\end{equation}
where $s^\mu$ the space-like spin polarization axis
\begin{equation}\label{smu}
s^\mu = k^\mu - {m^2 \over k\cdot q} q^\mu
\end{equation}
It is easy to show that the polarization tensors above satisfy the
normalization condition
\begin{equation}\label{norme}
e^{i}_\mu  e^{j,\mu}= {1\over 2}e^{i,\a \da} e^{i}_{\a \da}=
\delta_{ij} \ , \quad i,j=\pm,0
\end{equation}
and moreover satisfy the standard completeness relation for
massive spin one polarization vectors \cite{Spehler:1991yw}.

We use a standard notation for spinor variables as given in
appendix A. Polarization vectors for higher spins states can be
written by simple products of the above although to define
irreducible representations of the Lorentz $SO(1,3)$ group one
should impose the appropriate symmetries dictated by the given
Young Tableaux of the representation and of course transversality
and tracelessness. Transversality of the polarization tensors
(\ref{maspolar}) to the momentum vector of (\ref{momdec}) is
automatic. See \cite{Feng:2010yx} for an explicit example for the
massive spin 2 (level 2 state of the bosonic string).

\section{BCFW deformation for a 4-point scattering: General setup}\label{BCFWdef}
We want to study the $2\to 2$ scattering for open string tachyons
in bosonic string theory. We actually choose a configuration where
the tachyons carry only 4d momenta. In this manner the
intermediate states in the 4-point function will carry only 4d
momentum. We will deform particles 1 and 2. The deformation we
choose is
\begin{eqnarray}\label{momdef}
&&\hk_1= k_1+ n z \qquad \hk_2=k_2-n z \nonumber \\
&& k_i \cdot n= n\cdot n =0
\end{eqnarray}
and $n=\ket{2} \sbra{1}$. The masses are all taken to be $m^2_t=
-1/\apr$. Moreover we choose to decompose the other two momenta,
the unshifted ones, as
\begin{equation}\label{momdec34}
k_3= k_3^{\flat} + {m_t^2 \over s_{n3}^{\flat}} n\ , \qquad k_4=
k_4^{\flat} + {m_t^2 \over s_{n4}^{\flat}} n \ ,
\end{equation}
From the above, momentum conservation and (\ref{momdef}) it is
easy to see that
\begin{eqnarray}\label{nk34iden}
&&n\cdot k_3= n\cdot k_3^\flat \quad  n\cdot k_4= n\cdot k_4^\flat
\nonumber \\
&&n\cdot k_3^\flat=-n\cdot k_4^\flat
\end{eqnarray}
We also define the Mandelstam variables
\begin{equation}\label{mand}
s=(k_1+k_2)^2\ , \quad t=(k_1 + k_4)^2 \ , \quad u=(k_1 + k_3)^2
\end{equation}
It is very convenient to define the massless equivalents of the
Mandelstam variables
\begin{eqnarray}\label{mandless}
&&s^\flat= \braket{1 2}\sbraket{1 2}\ , \quad t^\flat= \braket{1
4}\sbraket{1 4}\ , \quad u^\flat= \braket{1 3}\sbraket{1 3}
\nonumber \\
&&\ts^\flat= \braket{3 4}\sbraket{3 4}\ , \quad \tit^\flat=
\braket{2 3}\sbraket{2 3}\ , \quad \tu^\flat= \braket{2
4}\sbraket{2 4}
\end{eqnarray}
Based on the decomposition above and momentum conservation one can
show that
\begin{eqnarray}\label{mandiden}
&&s^\flat= \ts^\flat \nonumber \\
&& s+t+u= 4 m^2_t \ , \quad s^\flat + t^\flat + u^\flat= - m^2_t \
, \quad s^\flat +\tit^\flat +\tu^\flat= - m^2_t
\end{eqnarray}
Finally the Mandelstam kinematic variables are given in terms of
the massless ones as
\begin{eqnarray}\label{mandconv}
&&s= s^\flat + {m^4_t \over s^\flat} + 2 m^2_t \nonumber \\
&& t= t^\flat + m^2_t( {\tu^\flat\over s^\flat} +2)=\tit^\flat +
m^2_t( {u^\flat\over s^\flat} +2)\nonumber \\
&& u= u^\flat + m^2_t( {\tit^\flat\over s^\flat} +2)=\tu^\flat +
m^2_t( {t^\flat\over s^\flat} +2)
\end{eqnarray}
Momentum conservation is written as follows
\begin{equation}\label{momcons}
\hk_1+\hk_2+k_3+k_4= \ket{1}\sbra{1} + \ket{2}\sbra{2}
+\ket{3}\sbra{3} +\ket{4}\sbra{4} + {m^2_t \over s^\flat}
(\ket{1}\sbra{1} + \ket{2}\sbra{2})=0 \ ,
\end{equation}
where we have used  (\ref{nk34iden}). Using the above we can
derive the useful identities
\begin{eqnarray}\label{spiniden}
&& {\braket{2 3} \over \braket{ 24}}= - {\sbraket{1 4} \over
\sbraket{ 1 3}} \nonumber \\
&& {\braket{2 1} \over \braket{ 2 4}} = - {\sbraket{34}\over
\sbraket{3 1}} + {m^2_t \over {\braket{24}\sbraket{12}}}
\end{eqnarray}
which of course in the limit $m^2_t \to 0$ go to the usual
identities for the scattering of four massless states.

Moreover since spinors live in a two dimensional space we can
always expand one of them as a linear combination of other two
i.e.
\begin{equation}
 - {\sbraket{34} \over \sbraket{31}}  \sbra{1} +
\sbra{4}= {\sbraket{1 4} \over \sbraket{1 3}} \sbra{3}
\end{equation}

\begin{figure}
\[{\cal A}_{4}^{(1,2)}\:=\:
 \raisebox{2.23cm}{\scalebox{.45}{{\includegraphics*[15pt,753pt][463pt,492pt]{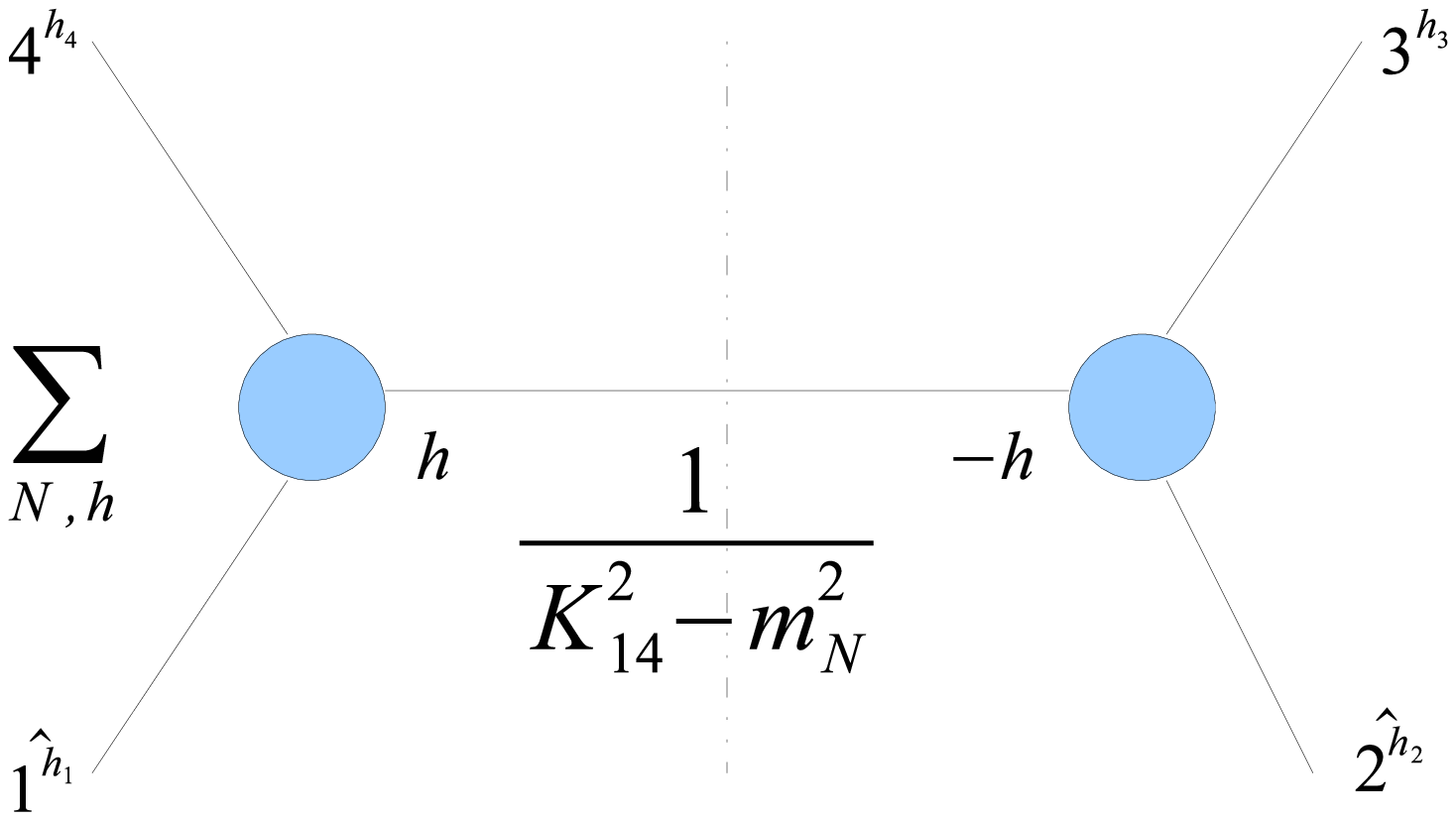}}}}\;+\;
 \raisebox{2.23cm}{\scalebox{.45}{{\includegraphics*[15pt,753pt][463pt,492pt]{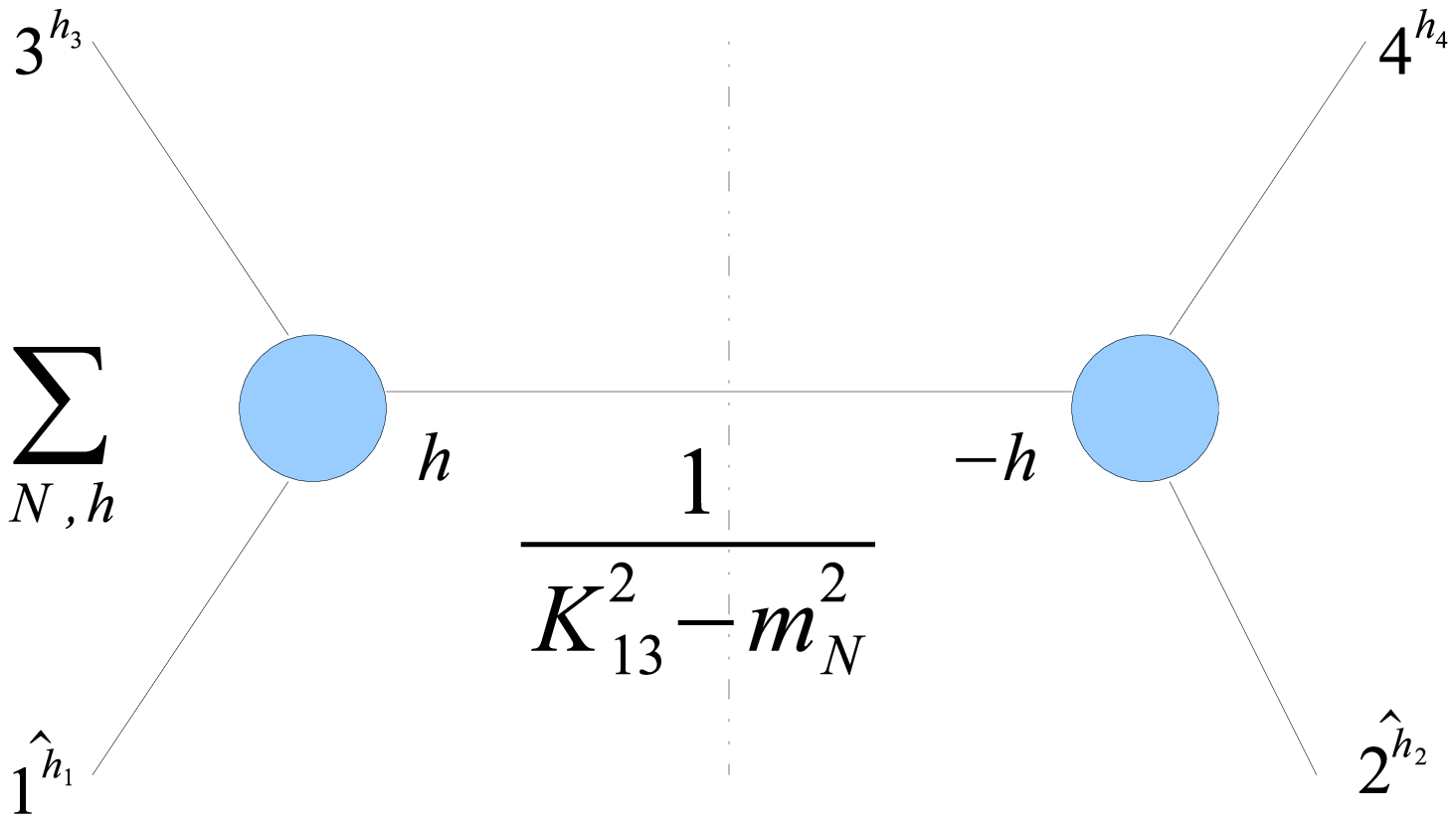}}}}
\]
\vspace{1cm} \caption{Factorization of a four-particle amplitude
into two on-shell three-particle amplitudes
\cite{Benincasa:2007xk},  where $K_{ij}^2= (k_i +k_j)^2$.
\label{BCFWfig}}
\end{figure}

We wish to employ the following BCFW recursion relation for the
4-point function. We need to calculate \cite{Benincasa:2007xk}
\begin{equation}\label{BCFWbasic}
{\cal A}^{(1,2)}_4(1,2,3,4) = \sum_{N,h(N)} \frac{{\cal
A}_{3}(\hat{1},4,\hat{k}_N)_{z_{res}} {\cal A}_{3}(-\hat{k}_N,
3,\hat{2})_{z_{res}}}{(\hk_1 + k_4)^2 - m_N^2} \ + \ (3
\leftrightarrow 4)
\end{equation}
The sum is over all string states of all levels $N$  and we sum
over their spins $h(N)$ with respect to the space-like spin axis
defined in (\ref{smu}). In figure \ref{BCFWfig} we show the two
terms which contribute above pictorially \cite{Benincasa:2007xk}.

For the time being we assume an abelian theory so the actual order
of the states in the 4- and 3- point function is irrelevant. When
we will calculate the non-abelian Veneziano amplitude we will be
more careful and define properly the order of states at the
various vertices.

From the relations above we derive, for the first term in
(\ref{BCFWbasic}) which we call the t-channel, that the poles on
the z-plane are given by the relation
\begin{equation}\label{t-res}
z_{res}= - { t-m_N^2 \over  2 n\cdot k_4}= -{ m^2_t -{m^2_N \over
2} + k_4^\flat \cdot k_1^\flat + {m_t^2 \over s_{12}^\flat}
k_4^\flat \cdot k_2^\flat \over  n\cdot k_4^\flat}
\end{equation}
and similar for the other term, which we will call the u-channel,
with $4\to 3$ and $t \to u$.

Now in order to implement the BCFW procedure we need to follow two
steps. First determine the polarization vectors of the
intermediate string state and second to apply this to the given
cubic vertex. We will discuss the cubic vertex in the following
section since this requires a more extended discussion and some
assumptions.

In order to define the polarization vectors (\ref{maspolar}) we
need to compute $\hk_N= \hk_1(z_{res}) +k_4$ and decompose it into
two light-cone vectors. We choose one of them to be $n$ since this
will simplify computations substantially later on. Moreover we
assume from now one that all hatted quantities are on the residue
$z_{res}$. The final result for the momentum vector is
\begin{eqnarray}\label{ksres}
&&\hk_N= \hk_N^\flat+  {m_N^2 \over 2 n\cdot k_4^\flat} n
=\ket{\hk_N}\sbra{\hk_N} + {m_N^2 \over
\braket{24}\sbraket{14}} \ket{2}\sbra{1} \nonumber \\
&& \ket{\hk_N}= {\sbraket{14} \over \sbraket{13}} \ket{4} + {m^2_t
\over \braket{12}\sbraket{13}} \ket{2} \quad \sbra{\hk_N}=\sbra{3}
- {m^2_t \over \braket{23}\sbraket{12}}\sbra{1}
\end{eqnarray}
The split of $\hk_N^\flat$ into the two spinors is ambiguous up to
phases $\a, \beta$ for $\ket{\hk_N}, \sbra{\hk_N}$ respectively
with $\a \cdot \beta=1$. If the spinor decompositions are done
correctly the dependance on these two phases should drop out in
the final result. Details of this computation appear in appendix
B. Using the relations above we can write the polarization vectors
\begin{eqnarray}\label{polres}
&&\has=\hk_N^\flat- {m_N^2\over 2 n\cdot k_N^\flat} n \, \quad
\he_0={\has \over m_N} \nonumber \\
&&\he^+= - \sqrt{2} {\ket{\hk_N} \sbra{1}\over \sbraket{31}} \quad
\he^-= - \sqrt{2} {\ket{2} \sbra{\hk_N} \over \braket{23}}
\end{eqnarray}
For completeness we also give the expressions
\begin{eqnarray}\label{hk12res}
&&\hk_1= k_1^\flat + {m^2_t\over s^\flat}k_2^\flat - {t-m^2_N\over
\braket{24}\sbraket{14}} \ket{2}\sbra{1} \nonumber \\
&&\hk_2= k_2^\flat + {m^2_t\over s^\flat}k_1^\flat + {t-m^2_N\over
\braket{24}\sbraket{14}} \ket{2}\sbra{1}
\end{eqnarray}

We will also need the products of polarization tensors with
momenta. Obviously by definition $\he^{\pm,0}\cdot \hk_N=0$. We
can easily derive the following
\begin{eqnarray}\label{hehk14}
&&\he^+\cdot \hk_1= -\he^+\cdot k_4= {m^2_t \over \sqrt{2}} \times
{\braket{23} \over \braket{12}\sbraket{13}}
\nonumber \\
&&\he^-\cdot \hk_1= -\he^-\cdot k_4= -{1\over \sqrt{2}}
{\braket{12}\sbraket{13} \over \braket{23}}\nonumber \\
&&\he^0\cdot \hk_1=-\he^0\cdot k_4= m_N
\end{eqnarray}

We also need the polarization products with the momenta $\hk_2$
and $k_3$
\begin{eqnarray}\label{hehk23}
&&\he^+\cdot \hk_2= -\he^+\cdot k_3= {1\over \sqrt{2}}
{\braket{23} \sbraket{12}\over \sbraket{13}}
\nonumber \\
&&\he^-\cdot \hk_2= -\he^-\cdot k_3= - {m^2_t  \over \sqrt{2} }
{\sbraket{13} \over \braket{23}\sbraket{12}}\nonumber \\
&&\he^0\cdot \hk_2=-\he^0\cdot k_3= -m_N
\end{eqnarray}
Notice that (\ref{hehk23}) are not given by a simple exchange of
labels between $1\to 2$ and $4\to 3$ in (\ref{hehk14}).

Using the above one can show the important equations
\begin{eqnarray}\label{polprod}
&&X^+=\he^- \cdot (\hk_1-k_4) \times \he^+ \cdot (k_3-\hk_2)=
2s^\flat
\nonumber \\
&&X^-=\he^+ \cdot (\hk_1-k_4) \times \he^- \cdot (k_3- \hk_2)=
2{m^4_t \over s^\flat}
\end{eqnarray}
and it is easy to show that $X^+ + X^-= 2(s-2m^2_t)$. In the
expressions above we have ordered the particles as it is dictated
by the cubic vertices of figure 2.

\section{Cubic vertices for the BCFW relations}\label{seccubic}
We know from \cite{Sagnotti:2010at} that the coupling of a leading
Regge trajectory state of level $N$ to two tachyons is given in
momentum space by the formula
\begin{equation}\label{leadregcoupCP}
{\cal A}_{0-0-N}(k_I, k_J, k_N)= \left({\cal A}^+_{bac}\ tr(\tau^a
\tau^b \tau^c) + {\cal A}^-_{abc}\ tr(\tau^b \tau^a \tau^c)\right)
\end{equation}
where $\tau^a$ are $SU(N)$ matrices in the adjoint and
\begin{equation}\label{leadregcouppm}
{\cal A}^{\pm; abc}(k_I,k_J,k_N)= {\cal J}_{N}^{\pm;ab }\cdot
\phi_N^c=(\pm\sqrt{\hapr})^N \ T_I^a \ T_J^b \ \phi_N^c\cdot
k_{12}^N \ .
\end{equation}
$T_I$ the two tachyon fields, $\phi_N$ the level N state of the
string, $k_{IJ}=k_I-k_J$ and ${\cal J}_{N}^{\pm;ab }\sim k_{12}^N
T_I^a T_J^b$ the Noether current built from two tachyons
\cite{Berends:1985xx,Sagnotti:2010at}. Notice that in the abelian
case where the Chan-Paton matrices become unit only even spin
particles couple to the two tachyon current. See also the
discussion in \cite{Benincasa:2007xk}.

This is a covariant coupling and applies to a leading Regge
trajectory state of string theory which is traceless and
transverse. We would like to infer the couplings of the subleading
Regge trajectories. This in principle is a rather complicated task
which would require full control of perturbative string theory.
Instead of constructing the vertex explicitly we can make an
educated guess. We conjecture the following cubic coupling in
string light-cone
\begin{equation}\label{cubcoupCPlc}
{\cal A}_{0-0-N}(k_I, k_J, k_N)= \left({\cal A}^+_{abc}(k_I, k_J,
k_N )\ tr(\tau^a \tau^b \tau^c) + {\cal A}^-_{bac}(k_I, k_J, k_N
)\ tr(\tau^b \tau^a \tau^c)\right)
\end{equation}

\begin{eqnarray}\label{cubcoupCPpm}
{\cal A}^{\pm; abc}&&(k_I, k_J, k_N ;H, h^+, h^-)= \nonumber\\
&&=(O_{abc})^{H}(\pm 1)^N(\sqrt{\hapr})^H \ T^a_I \ T^b_J \
{(e^+\cdot k_{IJ})^{h^+} \ (e^- \cdot k_{IJ})^{h^-}\over \big(
\prod_{n=1}^\infty (n^{h^+_n} h_n^+!)(n^{h^+_n}
h_n^+!)\big)^{1/2}}
\end{eqnarray}
where
\begin{eqnarray}\label{Hhdef}
&& h_n= h^+_n-h^-_n \ ,\ \ h=\sum_n h_n\ ,  \ \
N_n= h^+_n + h^-_n \nonumber \\
&& h^{\pm}= \sum_n h^{\pm}_n \ ,  \ H=\sum_{n} N_n\ ,\ \ N=\sum_n
N_n n
\end{eqnarray}
and the light-cone states are not constrained to be in any way
traceless but are naturally transverse. We define as $H$ the total
$e^{\pm}$ number or in other words the {\it highest spin} possible
for a given state, $h$ the total spin along the given axis $s^\mu$
and the subscripts $n$ are to distinguish the origin of each
$e^{\pm}$ in terms of the various levels of the string
oscillators. The function $O_{abc}$ is defined as follows for all
momenta incoming to the cubic vertex.
\begin{equation}\label{defO}
O_{abc}= \left( \begin{array}{cc} + & \qquad cyclic \
(a,b,c) \\
- &\qquad otherwise\end{array}\right)
\end{equation}
The purpose of this function is to give to the cubic amplitude the
right behavior under the exchange of the external states and in
particular under the exchange of the two tachyons. Its presence is
dictated by the symmetries of the covariant coupling for the
leading Regge trajectory as we will explain shortly.

We would like to justify this vertex. The key point is that we
claim that this is the coupling in string light cone in the
specific kinematic setup of BCFW (\ref{momdef},\ref{momdec34},
\ref{ksres}) which is nevertheless sufficient for the BCFW
procedure we wish to employ. In other words it is quite general in
order to allow using Lorentz invariance to recover the full
Veneziano amplitude for general kinematics. Lets give our main
arguments which actually sketches a possible rigorous proof which
we wont present here but leave for future work. A few more details
related to a possible light-cone String Field Theory derivation
are given in appendix C. The {\it natural} string light-cone frame
is defined in appendix B.

\begin{itemize}
\item {\bf The light-cone states and Regge trajectories.} Define
string oscillators which satisfy commutator relations
\begin{equation}\label{oscicom}
[\a^\mu_m, \a^\nu_n]= m\eta^{\mu \nu} \delta_{m+n} \ .
\end{equation}
As it is known in string light-cone the on-shell normalized
physical states are given solely in terms of the transverse level
$n$ oscillators $\a^i_{-n}$ through the expression
\begin{equation}\label{lcstate}
\ket{N, k} = \prod_{i=2}^{D-1} \ \prod_{n=1}^\infty
{(\a^i_{-n})^{N_{in}} \over \big(n^{N_{in}} N_{in}! \big)^{1/2}}
\ket{0;k}
\end{equation}
where $N=\sum_{in} N_{in} n$ the total level of the string state.
In \cite{Sagnotti:2010at} the $\braket{\phi_N, \phi_N}$ is the
symmetrized contraction of the space-time indices and the fields
are not normalized to unity while the states above are. Taking
this into account, we can write the leading Regge trajectory
covariant field $\phi_N$ using our normalized polarization vectors
(\ref{norme}) in four dimensions as
\begin{equation}\label{leadreggephi}
\phi_N = \sum_{h^{\pm}, h^0}^{ h^++h^-+h^0=N} {(\e^+)^{h^+}
(\e^-)^{h^-} (\e^0)^{h^0}\over (h^+!  h^-! h^0!)^{1\over2}}\ .
\end{equation}
This is the properly normalized field as in (\ref{lcstate}) for a
state of level $N$ built only from $a^{\pm,0}_{-1}$.

Moreover the tachyon current is totally symmetric and therefore as
noticed in \cite{Sagnotti:2010at} only the symmetric part of the
subleading Regge trajectories will couple. This means that the
subleading Regge states which couple to the current are built from
the symmetrized products of oscillators $a^{\pm,0}_{-n}$ of all
levels similar to (\ref{lcstate}).

\item {\bf The coupling to the current.} For the tree level
amplitudes we want to discuss we can confine ourselves
consistently to four dimensions. This is due to the fact that
since we have 4d external momenta also the Higher Spin (HS) or
reggeon state will have momentum on this 4d subspace. Moreover the
coupling above dictates that only polarizations vectors of the
reggeon state on this 4d subspace couple to the current built form
the two tachyon fields. In other words we do not have traces of
the reggeon state coupling to the current. In principle one could
have traces of the polarization tensor of the reggeon state in the
coupling (\ref{cubcoupCPpm}) which effectively would alter the
normalization coefficients. The dependance of the coupling in
terms of $N, H, h^+, h^-$ would be different and not the one
dictated by the assumption that all indices of a given light-cone
state couple to the tachyon current. The reasoning behind the
aforementioned  assumption lies into an appropriate choice string
light-cone frame.

For the purposes of discussing the light-cone setup, lets call the
light-cone frame momenta $k_I^{(\pm)}$ and the normal components
$k_I^i$. We use parentheses for $(\pm)$ in order  to avoid
confusion with the $e^{\pm}$ notation above, since this refers to
the normal coordinates $i=1,2$. As discussed in
\cite{Green:1987sp} in light-cone computations one cannot set all
$k^{(+)}_I=0$ therefore the vertex operators $e^{ik\cdot X}$
depend on the $X^{(-)}$ component in general. The $X^{(-)}$ in
light-cone quantization has an expansion which is quadratic in the
oscillators $a^i_{-n}$ and therefore if we compute correlators
with the states in (\ref{lcstate}) it results into traces. Also in
the operator language there are "chronic ordering" problems, as
explained in chapter 7 of \cite{Green:1987sp}, with the definition
of the $e^{ik^{(+)}X^{(-)}}$ operators, since $X^{(-)}\sim X^i
X^i$. So usually one has to use the method of light-cone string
field theory, see in example chapter 11 of \cite{Green:1987sp}.
Nevertheless for few external states it poses no problem to set at
least some of the $k^{(+)}_I$ momenta equal to zero while
preserving the mass-shell conditions $(k_I^i)^2=-m_N^2$. The later
assumes complex values for some $k^i_I$ but in any case we are
interested for the 3-point amplitude for complex momenta. We will
see in the next section that indeed the BCFW procedure for the
Veneziano amplitude can be employed consistently for CM energy
$s_{12}<0$ which implies complex momenta for $k_{1,2}$. In
appendix B we show that the massive momentum decomposition and
BCFW deformation we have employed in this setup leads to a natural
string light-cone frame. In this frame one of the tachyon external
states of the 3-point function has $\hk^{(+)}_1=0$. Moreover since
we have only one higher level state (reggeon), polarizations can
be chosen arbitrarily enough for our purposes to satisfy
transversality in this light-cone frame \footnote{In general
though with more higher level states in the 3-point function, this
might not be possible.}. The vectors $\he^{\pm}$ span the
transverse space to the string light-cone. With this setup it is
fairly easy to see that the coupling of the states (\ref{lcstate})
to the two tachyons should be proportional to $e^i\cdot k_{IJ}$ as
in (\ref{cubcoupCPpm}) and no traces of the polarizations should
appear \footnote{This can be done easily using the coherent state
oscillator formalism. We actually need to compute $$\bra{T_2} V_N
\ket{T_1} \ . $$ Pushing all annihilation oscillators of $V_N$ to
the right we do not meet any $a^{(-)}_{-n}\sim a^i_{-n} a^i_{-n}$
oscillators, since $k_1^{(+)}=0$, which would lead to traces in
this language.}. We call the transverse coordinates $i=\pm$ from
now on.

Notice that in the above we are referring to the 3-point function
we need for the BCFW and not the 4-point we wish to construct.
Obviously we assume that with a Lorentz boost we can take the
final BCFW constructed 4-point function to an arbitrary frame with
unrestricted kinematics. We also do not claim that the momenta of
all the n-external states can be put in the light-cone frame with
$k^{(+)}_I=0, \   I= 1, \dots n$. Of course our argument relies on
Lorentz invariance of the theory and the assumption of enough
generality to allow for the coupling in (\ref{cubcoupCPpm}) to
reproduce, via the BCFW procedure, the Veneziano amplitude. We
give a more explicit discussion on the construction of the vertex
based on light-cone String Field Theory in appendix C.

\item {\bf The phase factor $O_{abc}$.} We come now to the
$O_{abc}$ function which has an important role. First we will
present a formal argument based on cocycles for vertex operators
in string theory and then an argument based on the Lorentz
properties of the theory.

A formal argument is based on the following. In the operator
formalism computing a correlator one needs to be careful when
commuting the positions of two vertex operators. The vertex
operators of two tachyons pick-up co-cycle factors
\begin{equation}\label{cocycle}
e^{ik_1 \cdot X(x_1)} e^{ik_2 \cdot X(x_2)} \sim e^{i \pi 2\apr
k_1\cdot k_2 \epsilon(x_1-x_2)} e^{ik_2 \cdot X(x_2)} e^{ik_1
\cdot X(x_1)}
\end{equation}
where $\epsilon(x_1-x_2)= \pm 1$ for $x_1-x_2$ greater or less
than zero respectively. Using momentum conservation and mass-shell
conditions it is straightforward to derive that in a path integral
with a state of level N and the two tachyons we have
\begin{equation}\label{cocyclemashell}
2\apr k_1\cdot k_2= N+1 \ .
\end{equation}
Finally for a 3-point function all the operators have a Conformal
Killing Ghost $c(x_I)$ which needs to be commuted as well. Since
these are anticommuting fields they contribute and extra minus
sign. Therefore taking into account this and
(\ref{cocyclemashell}) in (\ref{cocycle}) we derive the final
phase $(-1)^N$ under the exchange of the two tachyons as expected.
It depends on the level of the massive string state and not on the
actual form of the vertex operator in terms of the oscillators.

We can actually see the presence of this factor in a more
intuitive manner through Lorentz invariance. For the abelian case
we know that only even spin couplings exist. The expression in
(\ref{cubcoupCPpm}) has the right property when inserted in
(\ref{cubcoupCPlc}). The $(-1)^N$ relative factor between ${\cal
A}_{\pm;abc}(I,J,N)$ shows that only even spin states contribute
which is at least correct for the leading Regge. In a covariant
coupling, once it is cast in terms of irreducible representations
of the massive little group, the remaining spin multiplets of a
given string level would have explicitly this property. Since we
are working in a light-cone frame some of this info is not so
obvious.

Take for example the spin 2 massive state. In light-cone the
possible states which can couple are composed of $a^i_{-1}
a^j_{-1}$ and $a^i_{-2}$ \footnote{The states $a^m_{-1}
a^\mu_{-1}$ and $a^\m_{-2}$ where $m= 4,\dots 26$ and $\mu=0\dots
26$ cannot couple due to the form of the two tachyon current in
(\ref{leadregcouppm}) for external tachyons with 4-dimensional
momenta. This holds for all the states of this class even with
traces by simple inspection of the string amplitude for the
3-point function of two tachyons and one massive string state.}.
The first state gives the spin $+2,-2,0$ components of the massive
spin 2 multiplet. We always refer to the spin component along the
axis of motion. The second state should give the spin $+1,-1$
states. If the coupling for this state is of the form in
(\ref{cubcoupCPpm}) without the function $O_{abc}$ this would lead
into a problem. An exchange of the two tachyons $I\leftrightarrow
J$ would give a minus the expression itself and this would mean
that it should vanish ((see also \cite{Benincasa:2007xk})). This
is obviously unacceptable since this state needs to couple to the
tachyon current due to Lorentz invariance of the theory. It is
needed to complete the massive spin 2 multiplet. The function
$O_{abc}$ takes care of this matter.

An easy way to see why it should be there is to consider how this
state emerges covariantly. The $a^{i}_2$ mode is the one of the
modes eaten via the Stuckelberg mechanism to give a mass to the
spin 2 multiplet. It corresponds to the mode $a^i_{-1} a^0_{-1}$.
Obviously this state is an even rank tensor and gives no problem
coupling to the tachyon current under the exchange of tachyon
states. One might wonder whether it is possible that one of the
subleading Regge trajectory states will couple only partially.
That is to couple only those states needed to complete the
multiplets coming from the higher spin states but leave the rest
out. The entangled way of Stuckelberg mechanism for mass
generation of the states and the high symmetry of the coupling
does not seem to leave any space for such a scenario. Of course
only a direct computation of the cubic coupling would eliminate
any doubt about it.

In the case with Chan-Paton factors we get the usual story of
(anti)commutator of Chan-Paton matrices for (even)odd level
states.

\item {\bf Connection to the covariant formulation.} Assume we
wanted to work in covariant formulation. How would the 3-point
amplitude compare to the one computed with the light-cone vertex?
Obviously they should have the same physical context modulo the
explicit appearance of massive little group representations of
$SO(3)$ for the covariant method. We have demonstrated that
$\he^0\cdot k_i \sim m_N= \sqrt{{N-1\over \apr}}$. So in our BCFW
relations, contractions of the zero component polarization tensor
give only spin and $\apr$ dependent terms. Therefore a state
$~(e^0)^{h^0}(e^+)^{h^+}(e^-)^{h^-}$ will give effectively a BCFW
cubic function: $c_N ({1\over
\apr})^{h_0/2}(e^+)^{h^+}(e^-)^{h^-}$, where $c_N$ a coefficient
dependent on the level of the state and its partitioning in the
three polarizations. Therefore in building the leading Regge
trajectory coupling for the various modes we will get indeed a
coefficient $(\sqrt{\hapr})^{N-h^0}= (\sqrt{\hapr})^{H}$, since
$N=H+h^0$. This is consistent with dimensional analysis of course.
The same holds for subleading Regge trajectories as well. The key
point is the normalization $c_N$ which we claim to be the one of
the light-cone states of string theory.

A direct contact with the covariant formulation would efficiently
proceed through the use of DDF \cite{Del Giudice:1971fp} states in
the OSFT \cite{Witten:1985cc} which can be shown to be equivalent
with the SFT in light-cone gauge \cite{Hata:1986kj}.

\item {\bf Completeness and Orthonormality.} We should emphasize
that application of the BCFW does not require to decompose the
intermediate state in the propagator in terms of irreducible
representations of the little group for massive states but rather
a complete and orthonormal set of states. The light-cone states
comprise such a set although they are not representations of the
massive states little group $SO(3)$ in 4d but rather of the
massless one $SO(2)$.

\end{itemize}

After the above discussion we can rewrite in a more manageable
form the couplings with CP factors
\begin{eqnarray}\label{cubcoupCPlcfinal}
{\cal A}_{0-0-N}(k_I, k_J, k_N)&=& (O_{abc})^{H}(\sqrt{\hapr})^H \
T^a_I \ T^b_J \nonumber \\
&&{(e^+\cdot k_{IJ})^{h^+} \ (e^- \cdot k_{IJ})^{h^-}\over \big(
\prod_{n=1}^\infty (n^{h^+_n} h_n^+!)(n^{h^+_n}
h_n^+!)\big)^{1/2}}\ tr([\tau^a, \tau^b]_{\pm} \tau^c)
\end{eqnarray}
where the $[\cdot  \ ,\cdot]_\pm$ means (anti)commutator and
applies for (even)odd level respectively.

Based on the 3-point amplitude above we will compute the BCFW
equation (\ref{BCFWbasic}). We conclude this section by pointing
out that from the ansatz above the main assumptions were the
actual normalization of the 3-point function for each light-cone
state and the choice of string light-cone frame  which sets some
of the $k^+_I=0$. Both the form of the vertex, based on the
tachyon current, and the sign factor $O_{abc}$ are pretty much
constrained by the symmetries of the vertex and Lorentz invariance
of the theory. The power of $\apr$ is dictated by dimensional
analysis. In appendix C we attempt an non-rigorous light-cone SFT
derivation which gives nevertheless the full result in
(\ref{cubcoupCPpm}) even up to the correct normalization. A
light-cone or even covariant perturbation theory proof would be
highly desirable.

\section{3-point functions and combinatorics}\label{combin}
The residues of (\ref{BCFWbasic}) are products of two 3-point
functions ${\cal A}_L$ and ${\cal A}_R$. Each $\he^+$ on ${\cal
A}_L$ should be accompanied by an $\he^-$ on ${\cal A}_R$ and
vice-versa since the intermediate state has $+h$ spin for the one
amplitude and $-h$ for the other. Moreover the orthogonality of
states (\ref{lcstate}) is based on the orthogonality of the
corresponding oscillators upon which they are built. This means
that the state which appears on ${\cal A}_R$ should be the
conjugate of the one in ${\cal A}_L$ i.e. $(\he^{\pm})^* =
\he^{\mp}$. So Based on (\ref{cubcoupCPlcfinal}) and
(\ref{polprod}) we see that, each oscillator with a given fixed
occupation number $N_n$ in a given state gives a contribution to
the product of the two 3-point functions as follows
\begin{eqnarray}\label{Nncontr}
&&({\cal A}_L \times {\cal A}_R)_{N_n}= \sum_{h^+_n}
(\hapr)^{h^+_n + h^-_n} {(X^+)^{h^+_n}(X^-)^{h^-_n} \over h^+_n!
h^-_n!
n^{h^+_n+h^-_n}}=\nonumber \\
&&=\sum_{h^+_n} {1\over N_n!} \ \bn{N_n}{h^+_n} { (\hapr
X^+)^{h^+_n}(\hapr X^-)^{h^-_n}\over n^{N_n}}=
\nonumber \\
&&{1\over N_n!}{(\a(s)+1)^{N_n}\over n^{N_n}}
\end{eqnarray}
where the summation is over all partitionings of $N_n$ in $h^+_n$
with the condition $N_n=h^+_n+h^-_n$. We also have defined the
Regge trajectory
\begin{equation}\label{reggesl}
\a(s)=\apr s +1
\end{equation}
Now we are in position to recover the full contribution of a state
of (\ref{lcstate}) with a given total highest spin $H$
\begin{equation}\label{Hcontr}
({\cal A}_L \times {\cal A}_R)_{H} = \sum_{part N_n} \ \prod_{N_n}
\ {1\over N_n!}{(\a(s)+1)^{N_n}\over n^{N_n}}=\sum_{part N_n} \
\prod_{N_n} {N! \over N_n! n^{N_n}} \ {(\a(s)+1)^{H}\over N!}
\end{equation}
where the summation is over all partitionings of the total
polarization number (or highest spin $H=\sum_n N_n$) into the
integer numbers $N_n$. In the last equation we have defined
$N=\sum_n N_n n$ which is actually the level of the given state.

It turns out that the coefficients of $(\a(s)+1)^H/N!$ in the
expression in (\ref{Hcontr}) are the definition of the Stirling
numbers of the first kind
\begin{equation}\label{Stirling}
S(N,H)= \sum_{part N_n} \ \prod_{N_n} {N! \over N_n! n^{N_n}}\ ,
\quad N= \sum_n N_n n \ , \ \ H= \sum_n N_n
\end{equation}
These count the number of permutations of N elements with H
disjoint cycles \cite{stirling}.

To compute the total contribution of a given level N state we need
to sum over all the possible oscillator configurations of total
number $H$ with upper limit $N$. Actually the Stirling numbers of
the first kind generate the Pochhammer symbols and it turns out
that
\begin{equation}\label{poch}
({\cal A}_L \times {\cal A}_R)_{N}={1\over N!} \sum_{H=0}^N
S(N,H)(\a(s)+1)^H= {1\over N!} (\a(s)+1)_N= {\Gamma (\a(s)+1+N)
\over N! \ \Gamma (\a(s)+1)}
\end{equation}
In appendix D we show a few examples for the first few levels.
Lets consider summing up the result above over all levels in
(\ref{BCFWbasic}). This is not something we do for the abelian
case, where only even levels contribute, but certainly it is
instructive. The naive result is given by
\begin{eqnarray}\label{BCFWbeta}
{\cal A}^{(1,2)}_{4}(1,2,3,4 &;& t-channel)= \sum_{N=0}^{\infty}\
{1\over N!}
{(\a(s)+1)_N \over \a(t)-N}= \\
&&\sum_{N=0}^{\infty}\ {1\over N!} {(\a(s)+1)(\a(s)+2)\dots
(\a(s)+N) \over \a(t)-N}= -B (-\a(s), -\a(t))\nonumber
\end{eqnarray}
We see that this is one of the terms of the Veneziano amplitude.
The series above is defined for the region where $Re(\a(s)) <0$
and indeed the Pomeron analysis \cite{Cheung:2010vn, Boels:2010bv,
Fotopoulos:2010cm} shows that BCFW relations should hold in this
region since the asymptotic behavior of the string amplitude under
BCFW tachyon deformation is expected to be $z^{1+\apr s}$.
Actually our expansion should be valid in the more constrained
domain $|\a(s)+1| < 1$. We are summing perturbatively over
vertices with increasing power of momenta. We can see that this is
the case for our result by noticing that $\a(s)+1= \apr k_1\cdot
k_2$. Therefore indeed the actual expansion parameter is $\a(s)+1$
and should be in the unit circle on the complex plane for the
series to converge.

Then the series can be extended analytically on the whole complex
$s$-plane. This in turn demonstrates beautifully the dual nature
of the amplitude where the s-channel poles appear after resuming
over all the t-channel poles \cite{Green:1987sp}. In the next
section we will show explicitly how the results above apply to the
Veneziano amplitude for the Abelian and non-Abelian cases.

\section{Veneziano amplitude via BCFW}\label{VenBCFW}
Lets apply (\ref{BCFWbasic}) to the abelian case using
(\ref{poch}) for the contribution of each string level. As we have
emphasized before only even level states contribute so we do not
expect to derive a beta function immediately as in
(\ref{BCFWbeta}). The two terms in (\ref{BCFWbasic}) give
\begin{equation}\label{Venabelian}
{\cal A}^{(1,2)}_4(1,2,3,4)= B^{e}(-\a(s);-\a(t)) +
B^{e}(-\a(s);-\a(u))
\end{equation}
where we have defined the following expressions
\begin{eqnarray}\label{Beo}
&&B^e(-x;-y)=\sum_{N \in 2 \mathbb{N}}\ {1\over N!} {(x+1)_N \over
y-N}\nonumber \\
&&B^o(-x;-y)=\sum_{N \in 2 \mathbb{N}+1}\ {1\over N!} {(x+1)_N
\over y-N}
\end{eqnarray}
The expressions above do not have the appropriate duality
properties of the beta functions so the order of their arguments
is important. They can be rewritten in terms of hypergeometric
functions but there is no obvious way to transform them into beta
functions as expected.

Now we will show the expression in (\ref{Venabelian}) is actually
proportional to the abelian Veneziano amplitude
\begin{equation}\label{Ven}
{\cal A}_4=-{1\over 2} \left[ B (-\a(s), -\a(t))+B (-\a(s),
-\a(u))+B (-\a(t), -\a(u))\right]
\end{equation}

The first and most important observation is that the two results
in (\ref{Venabelian}) and (\ref{Ven}) have the same residues in
the t- and u-channels. The odd residues of the beta function in
the t-channel cancel between $B (-\a(s), -\a(t))$ and $B (-\a(t),
-\a(u))$ and similar statement holds for the odd u-channel poles.
Now it is obvious that the result in (\ref{Venabelian}) should
reproduce the $B (-\a(t), -\a(u))$ term although it is not evident
at first sight. We use the exact same reasoning as in
\cite{Green:1987sp}) chapter 1 for the beta function written as a
series of its poles in one channel in the region where in other
channel it has no pole.

In the region of series convergence $Re(\a(s))<0$, (\ref{Ven}) has
no s-channel poles. Moreover the residues of the two expressions
(\ref{Ven}) and (\ref{Venabelian}) agree. Therefore they can only
differ by an entire function in the $t,u$ variables. Since such an
entire function which vanishes for large $|\a(t)|$ and $|\a(u)|$
does not exist then we can conclude that the two expressions are
identical. Therefore we just need to show that the Veneziano
amplitude (\ref{Ven}) vanishes when either $|\a(t)|$ or $|\a(u)|$
go to infinity. Actually it turns out that both moduli of these
variables need to go to infinity at the same time and not just
either of them. The reason is momentum conservation
\begin{equation}\label{momconsVen}
\a(s)+ \a(t) +\a(u)=-1
\end{equation}
In the previous section we pointed out that actually the BCFW
procedure we employed is valid only in the more constrained domain
$-2<\a(s)<1$. Therefore in the large $|\a(t)|$ regime we have
$\a(u)\simeq -\a(t)$ and $|\a(u)|$ goes so infinity as well.
Applying this limit to (\ref{Ven}) and using Stirling
approximation we get
\begin{equation}\label{Venasym}
{\cal A}_4 \simeq \Gamma(-\a(s)) \left( (-\a(t))^{\a(s)} +
(-\a(u))^{\a(s)}\right) + (\a(t))^{-{1 \over  2}} \
2^{\a(s)-{3\over 2}}
\end{equation}
We see that indeed for $Re(\a(s))<0$ the amplitude goes to zero as
expected due to the constructibility proof using Pomeron
operators. We can analytically continue this expression outside
the region of convergence of the series and this we way we will
recover the full Veneziano amplitude. This concludes the proof of
the BCFW construction for the abelian Veneziano amplitude. One can
check the constructibility criterion of \cite{Benincasa:2007xk} by
deforming the particles 1 and 4. The final result is the same.

Lets summarize the following identities which are derived with the
methodology above based on the assumptions of convergence of the
series and also momentum conservation (\ref{momconsVen}). They
will be useful for the rest of this section.
\begin{eqnarray}\label{Betaiden}
&& B^{e}(-x;-y) + B^{o}(-x;-y)= -B(-x,-y) \nonumber \\
&& B^{e}(-\a(s);-\a(t)) + B^{e}(-\a(s);-\a(u))=\nonumber \\
&&=-{1\over 2} \left( B (-\a(s), -\a(t))+B (-\a(s), -\a(u))+B
(-\a(t), -\a(u))\right)
\nonumber \\
&& B^{o}(-\a(s);-\a(t)) + B^{o}(-\a(s);-\a(u))=\nonumber
\\
&&=-{1\over 2} \left( B (-\a(s), -\a(t))+B (-\a(s), -\a(u))-B
(-\a(t), -\a(u))\right)
\end{eqnarray}

Now we will proceed with the Non-abelian case. Using the vertices
in (\ref{cubcoupCPlcfinal}) it is straightforward to derive from
the two terms in (\ref{BCFWbasic}) the following
\begin{eqnarray}\label{Vennonabe}
&&{\cal A}_{4}^{(1,2)} = {\cal A}^{(1,2)}_t + {\cal A}^{(1,2)}_u = \\
&&=tr \left( \{t^1,t^4\} \{t^3,t^2\}\right)B^e (-\a(s);-\a(t)) +
tr \left( [t^1,t^4] [t^3,t^2]\right)(B^o (-\a(s);-\a(t))) +
\nonumber \\
&&+tr \left( \{t^1,t^3\} \{t^4,t^2\}\right)B^e (-\a(s);-\a(u)) +
tr \left( [t^1,t^3] [t^4,t^2]\right)(B^o (-\a(s);-\a(u)))\nonumber
\end{eqnarray}
where we have used the standard notation for (anti)commutators.
Expanding in color ordered factors we get
\begin{eqnarray}\label{VencolorBCFW}
&&{\cal A}_{4}^{(1,2)}= tr\left( t^1 t^4 t^3 t^2 + t^1 t^2 t^3
t^4\right) \left(B^{e}(-\a(s);-\a(t))+
B^{o}(-\a(s);-\a(t))\right)+
\nonumber \\
&&+tr\left( t^1 t^3 t^4 t^2 + t^1 t^2 t^4 t^3\right)
\left(B^{e}(-\a(s);-\a(u))+ B^{o}(-\a(s);-\a(u))\right)+
\nonumber \\
&&+tr\left( t^1 t^4 t^2 t^3 + t^1 t^3 t^2 t^4\right)
(B^{e}(-\a(s);-\a(t))+B^{e}(-\a(s);-\a(u))- \nonumber \\
&&-B^{o}(-\a(s);-\a(t))-B^{o}(-\a(s);-\a(u)))
\end{eqnarray}
and using the identities in (\ref{Betaiden}) we easily derive the
Veneziano amplitude for the non abelian case
\begin{eqnarray}\label{Vencolor}
&&{\cal A}_{4}= -tr\left( t^1 t^4 t^3 t^2 + t^1 t^2 t^3 t^4\right)
B(-\a(s);-\a(t))-
\nonumber \\
&&-tr\left( t^1 t^3 t^4 t^2 + t^1 t^2 t^4 t^3\right)
B(-\a(s);-\a(u))-
\nonumber \\
&&-tr\left( t^1 t^4 t^2 t^3 + t^1 t^3 t^2 t^4\right)
B(-\a(t);-\a(u))
\end{eqnarray}
A deformation of the particles 1 and 4 gives identical result as
required by the consistency condition of \cite{Benincasa:2007xk}.

\section{Conclusions and Outlook}\label{conclusions}
We have proposed an explicit construction of the Veneziano
amplitude based on the BCFW procedure. We first applied the
methodology of spinor helicity formulation for massive states as
i.e. \cite{Boels:2010bv}. Then we proceeded with a conjecture on
the 3-point function of an arbitrary massive state of string
theory with two tachyons in a kinematic frame we consider general
enough for our purposes. We gave some arguments which support its
form but a lack of rigorous proof is definitely an unsatisfying
state of affairs. We applied the conjectured 3-point function to
the BCFW recursion formulas for the construction of the 4-point
function for four external tachyons. After some interesting
combinatorics we managed to show how the celebrated Beta functions
of the Veneziano amplitude arise. We sum up an infinite number of
massive string exchanges in the channels which become deformed
under the BCFW deformation. The dual behavior of the Veneziano
amplitude emerges from an analytic continuation of the formal
series in the kinematic region where poles in the dual channels
appear.

Thinking inversely our cubic couplings can be taken as derived via
the BCFW method. The residues of the BCFW deformation of the
4-point function determine the product (\ref{Hcontr}) which can be
reproduced as we have shown by the cubic couplings
(\ref{cubcoupCPlcfinal}). So our proposed 3-point function can be
taken as derived rather than conjectured although the inverse
procedure we claim above does non necessarily lead to unique
3-point functions (see for example \cite{Cheung:2010vn} where the
BCFW deformation of the five tachyon amplitude can be written as a
recursion relation in terms of only tachyon sub-amplitudes).

We would like to make a few comments and clarifications regarding
the BCFW procedure itself for string amplitudes. One should not
confuse the BCFW recursion relations with the usual string
world-sheet factorizations when a kinematic invariant approaches a
physical pole \footnote{A related discussion regarding string
factorizations appeared in an updated version of
\cite{Boels:2010bv} when the present work was in its final
details.}. Definitely due to unitarity the string amplitude
factorizes on the poles to lower on-shell string amplitudes. But
these amplitudes are defined on the special kinematics of the
external states and do not constitute a recursive construction of
the string amplitude away from these special points. In the BCFW
relations the lower point amplitudes are computed on complex
momenta which differ from the momenta of the actual physical
residues.

Moreover the BCFW method reconstructs the string amplitude using
the residues of only a subset of all the possible channels. These
are the Reggerized channels. In the Pomeron language used in
\cite{Cheung:2010vn,Boels:2010bv,Fotopoulos:2010cm} these channels
correspond to the shifted kinematic variables. The large Regge
behavior is dictated by the unshifted kinematic variable i.e.
${\cal A}\sim(\hat{t})^{\apr s}$. Keeping the unshifted kinematic
variables in the region where the Regge channel becomes damped for
large shifts, we are guaranteed the absence of a residue at
infinity. The un-reggerized channels are reconstructed by analytic
continuation of the unshifted momenta outside the aforementioned
region. The BCFW reconstruction of the string amplitude looks like
an expansion in terms Feynman diagrams for tree level  field
theory exchanges. In the region where the BCFW method is
applicable we can think of the world-sheet as thinning out as we
approach the poles of the deformed amplitude. The residues of the
poles in the Reggerized channel reconstruct in a way the
world-sheet locally. Then moving outside this region we recover
the full world-sheet. It is in complex momenta for the exchanged
particle that this description is possible without restricting the
momenta of the external states to special values.

Another thing to point out is that, just as in gauge theory
applications of BCFW, individual terms of the recursion relations
might exhibit spurious poles (see i.e. \cite{ArkaniHamed:2009dn})
which do not correspond to physical poles. These should cancel in
the final expression and indeed this is a strong consistency check
for the validity of the BCFW method.

There are several directions one should follow. The foremost
important is a rigorous derivation of the 3-point amplitude
proposed in (\ref{cubcoupCPlcfinal}). Definitely if such a result
cannot be reconstructed then it will be quite remarkable that the
conjectured 3-point vertex reproduces the Veneziano amplitude. It
is not clear what would be the physical meaning of such a result.

The next step would be to use the general 3-point vertex for
external string states of arbitrary level (reggeons)
\cite{Ademollo:1974kz} with appropriate choices of kinematic
regions to see if we can reproduce higher point tachyon
amplitudes. Even for recursion relations for the five tachyon
amplitude one would need at least the one tachyon with two
reggeons amplitude. This way one could compute the 4-point
function of three tachyons and on reggeon which is needed for
applying the BCFW method in this case. Constructing the five
tachyon amplitude is highly non-trivial task which will uncover
any potential problems in applying the BCFW method for string
amplitudes.

Obviously the most interesting cases lie in the context of
superstring theory. Tree level gluon amplitudes do not depend on
the compactification details of the internal coordinates in string
theory \cite{Stieberger:2006bh}. So it is very interesting
phenomenologically to construct recursion relations which allow us
to compute the leading string corrections to the n-gluon
amplitudes. Therefore an obvious generalization of the present
work is its extension to superstring theory. A good starting point
would be the supersymmetric three reggeon vertex of
\cite{Hornfeck:1987wt}.

Beyond the obvious applications for string amplitude computations
one of the motivations for the present work is to try to extend to
string theory many of the present developments, like i.e. the
Grassmannian \cite{ArkaniHamed:2009dn}, integrability
\cite{Beisert:2006ez} for ${\cal N}=4$ SYM, BCFW relations for
loop amplitudes \cite{ArkaniHamed:2010kv,Boels:2010nw} e.t.c. In
this case we might be able to learn more about the symmetries of
the theory or its non-perturbative properties.

\acknowledgments We would like to thank T.~Taylor and R.~Boels for
valuable comments and in particular C.~Angelantonj and N.~Prezas
for stimulating and insightful discussions. In addition N.~Prezas
for comments and corrections on the manuscript prior to
publication and ongoing discussions on related subjects. The work
of A.~F. was supported by an INFN postdoctoral fellowship and
partly supported by the Italian MIUR-PRIN contract 20075ATT78.

\renewcommand{\thesection}{A}
\setcounter{equation}{0}
\renewcommand{\theequation}{A.\arabic{equation}}
\section*{Appendix A: Notation and a Basis of Vectors}\label{A}
We use the definitions
\begin{equation}\label{bra}
\braket{\la \la '}= \epsilon^{\a \b} \la_{\a} \la_{\b}'\ ,  \quad
\sbraket{\tla \tla '}= \epsilon^{\a \b} \tla_{\da} \tla_{\dbe}'
\end{equation}
and for light-cone vectors we use
\begin{equation}\label{momsig}
k^\mu= \la^{\a} \sigma^\mu_{\a \da} \tla^{\da}\ , \quad 2 p\cdot
q= \braket{\la ^p \la ^q}\sbraket{\tla ^p \tla^q}
\end{equation}
The metric we use is $(+,-,-,-)$ and $\sigma^\mu=(1, \sigma^i)$.
The following identities and definitions are used in many cases in
the spinor helicity formalism \begin{eqnarray}
&&(\bar{\sigma}^\mu)^{\a \da}=-\epsilon^{\a \beta} \epsilon^{\da
\dbe} (\sigma^\mu)_{\beta \dbe}, \quad \epsilon^{12}=
\epsilon_{12}=1\nonumber \\
&&(\bar{\sigma}_\mu)^{\a \da}(\sigma^\mu)^{\beta \dbe}= 2
\delta^\beta_{\a} \delta^{\da}_{\dbe} \ .
\end{eqnarray}
With the definitions above we can write an arbitrary 4-vector
$k^\mu$ in matrix notation
\begin{equation}\label{kmatrix}
k^\mu = \left(\begin{array}{cc} k^0+k^3 \ & k^1 + i k^2
\\ k^1-ik^2 \ & k^0-k^3
\end{array}\right)
\end{equation}
where $\det(k^\mu)=m^2$. Polarization for integer spin-$s$
massless particles are given in terms of the polarizations for
massless spin 1 particles
\begin{equation}
e^+_{\a_1 \da_1, \dots \a_s \da_s}=\prod_{i=1}^s e^+_{\a_i \da_i}\
, \quad e^-_{\a_1 \da_1, \dots \a_s \da_s}=\prod_{i=1}^s e^-_{\a_i
\da_i}
\end{equation}
where the spin 1 polarization vectors are given by
\begin{equation}\label{polarphoton}
e^+_{\a \da}=  \sqrt{2}\ { \la_{\a} \tilde{\mu}_{\da} \over
\sbraket{\tilde{\mu} \la} } \ ,\quad e^-_{\a \da}= \sqrt{2} \ {
\mu_{\a} \tla_{\da} \over \braket{\mu \tla} }
\end{equation}
and $\mu_{\a}$, $\tilde{\mu}_{\da}$ arbitrary reference spinors. A
change in the reference spinors corresponds to a gauge
transformation of the polarization tensors for the massless states
and a given amplitude must be invariant under such a
transformation.

\renewcommand{\thesection}{B}
\setcounter{equation}{0}
\renewcommand{\theequation}{B.\arabic{equation}}
\section*{Appendix B: Polarization calculation}\label{B}
The first relation we will prove is (\ref{ksres}). Write the
momentum of the intermediate state using
\begin{equation}\label{Bhk}
\hk_N= k_1^\flat + k_4^\flat + {m^2_t \over s^\flat} k_2^\flat +
{m^2_t \over 2k_4^\flat \cdot n} n - {m^2_t -m^2_N/2 + k_4^\flat
\cdot k_1^\flat + {m^2_t \over s^\flat} k_4^\flat \cdot k_2^\flat
\over n \cdot k_4^\flat} n
\end{equation}
We can show the following identities:
\begin{equation}\label{Biden1}
k_1^\flat + k_4^\flat - n {k_4^\flat \cdot k_1^\flat \over
k_4^\flat \cdot n}= {\braket{21}\over \braket{24}} \ket{4}
\sbra{1} + \ket{4}\sbra{4}
\end{equation}
using (\ref{spiniden}) and linear dependance of the spinors
$\sbra{1}, \sbra{3} , \sbra{4}$.
\begin{equation}\label{Biden2}
{m^2_t\over s^\flat} (k_2^\flat - {k_4^\flat\cdot k_2^\flat\over
k_4^\flat \cdot n} n)= {m^2_t \over \braket{12}\sbraket{14}}
\ket{2}\sbra{4}
\end{equation}
using linear dependance of $\sbra{1}, \sbra{2} , \sbra{4}$.
\begin{equation}\label{Biden3}
{m^2_t \over \braket{12}\sbraket{14}} \ket{2}\sbra{4}- {m^2_t
\over \braket{24}\sbraket{14}} \ket{2}\sbra{1}= {m^2_t \over
\braket{12}\sbraket{13}} \ket{2}\sbra{3}+ {m^4_t \over s^\flat
\braket{24}\sbraket{14}} \ket{2}\sbra{1}
\end{equation}
using the second of (\ref{spiniden}) and linear dependance of
$\sbra{1}, \sbra{3} , \sbra{4}$.
All of the above lead to (\ref{ksres}).

From the polarization contractions only the $\he^0 \cdot \hk_1
=-\he^0 \cdot k_4$ is a tedious one. The other ones come easily
with application of (\ref{spiniden}). We show some intermediate
steps form the $\he^0$ polarization manipulations
\begin{eqnarray}\label{Biden4}
&&\he^0 \cdot \hk_1= {\hk_N^\flat \cdot \hk_1 \over m_N}=
\nonumber \\
&&={-1\over 2m_N} \{ {\sbraket{14}\over \sbraket{13} }(
\braket{41}\sbraket{13} + m^2_t {\braket{42}\sbraket{23}\over
s^\flat}) + (t-m_N^2)+ {m^4_t\over s^\flat} - m^2_t \}= \nonumber
\\
&&={-1\over 2m_N}\{ -t^\flat + m^2_t ({\tit^\flat\over s^\flat}-1)
+{m^4_t \over s^\flat} + (t-m^2_N) \}= \nonumber \\
&&= {m_N\over 2}
\end{eqnarray}
One can easily guess that $\he^0 \cdot \hk_1 =-\he^0 \cdot k_4$
since $\he^0 \cdot \hk_N=0$ but as a useful cross-check of our
calculations we can explicitly show that
\begin{eqnarray}\label{Biden5}
&&\he^0 \cdot k_4= ( \hk_N^\flat - {m^2_N \over 2 n \cdot
\hk_N^\flat} n) \cdot {k_4 \over m_N}= \nonumber \\
&&=-{1\over 2 m_N} \{ ( -m^2_t + m^2_t
{\braket{24}\sbraket{43}\over \braket{12}\sbraket{13}} - {m^4_t
\over s^\flat}) + m^2_N\}
=\nonumber \\
&&=-{m_N\over 2}
\end{eqnarray}

Now we proceed in defining a convenient "light-cone" basis of
vectors which will turn useful in discussing the 3-point amplitude
needed for the BCFW relations. We use the light-cone vectors used
to define $\hk_N$ in order to write down a basis as in
\cite{Boels:2010mj}
\begin{eqnarray}\label{kNbasis}
&&v_1= {\ket{\hk_N}\sbra{\hk_N}\over m_N} \ , \quad  v_3=
{\ket{2}\sbra{\hk_N}\over \braket{2 \hk_N}} \nonumber
\\
&& v_2= m_N {\ket{2}\sbra{1}\over \braket{2
\hk_N}\sbraket{1\hk_N}} \ , \quad v_4={\ket{\hk_N}\sbra{1}\over
\sbraket{1 \hk_N}}
\end{eqnarray}
The elements $v_1$ and $v_2$ are the light-cone vectors used to
define the momentum vector $\hk_N$ and the other two vectors $v_4$
and $v_3$ are proportional to the polarization vectors $\he^+$ and
$\he^-$ respectively. There exists a frame where the basis vectors
in (\ref{kNbasis}) take the form
\begin{equation}\label{eq:choiceofframe}
\begin{array}{cccc} v^1_{\mu} \sim  \left(\begin{array}{c} 1 \\ 0 \\ 0 \\ 1 \end{array}\right) \quad &
v^2_{\mu} \sim  \left(\begin{array}{c} 1 \\ 0 \\ 0 \\ -1
\end{array} \right) \quad  & v^3_{\mu} \sim \left(\begin{array}{c}
0 \\ 1 \\ i \\ 0 \end{array}\right) \quad  & v^4_{\mu} \sim
\left(\begin{array}{c} 0 \\ 1 \\ -i \\ 0 \end{array} \right)
\end{array}
\end{equation}
with \begin{equation}\label{viden} v^{1;\mu} v^2_\mu + v^{3;\mu}
v^4_\mu=0.\end{equation} As explained in \cite{Boels:2010mj} an
arbitrary 4-vector can be expanded in the following manner
\begin{eqnarray}\label{momexpansion}
&&k_\mu= c_1 v_1+ c_2 v_2+ c_3 v_3 + c_4 v_4= \nonumber \\
&&\equiv k_\mu^\flat + {k^2\over 2 v_3 \cdot k} v_3
\end{eqnarray}
where the basis vector $v_3$ was chosen arbitrarily to define the
second light-cone vector, in addition to $k^\flat_\mu$, needed for
the decomposition of the vector $k_\mu$. We can determine the
coefficients $c_i$ as follows
\begin{eqnarray}\label{ci}
&& c_1=-m_N {\bra{2}k\sket{1} \over 2 \hk^\flat_N \cdot n} \ ,
\quad c_2=-{\bra{\hk_N}k\sket{\hk_N}\over m_N}
\nonumber \\
&& c_3=-{\bra{\hk_N}k\sket{1} \over \sbraket{1\hk_N}} \ , \quad
c_4=-{\bra{2}k\sket{\hk_N} \over  \braket{2\hk_N}}
\end{eqnarray}
We can indeed check that for an arbitrary vector with
decomposition
$$ k^\mu= a^\mu + {m^2 \over \braket{ab}\sbraket{ab}} b^\mu$$
the determinant of the matrix
$$
\left(\begin{array}{cc} c_1 & c_4
\\ c_3 & c_2
\end{array}\right)$$ is equal to $m^2$ as it is required by
(\ref{kmatrix}) using (\ref{viden}). To prove the statement above
we have to use the Schouten identity
\begin{equation}\braket{xy}\braket{wz} + \braket{xz}\braket{yw}+
\braket{xw}\braket{zy}=0 \ .
\end{equation}
Then it is easy to see that in this frame the momentum vectors of
the three particles $\hk_1, \ k_4$ and $\hk_N$ are written in
matrix notation as
\begin{eqnarray}\label{mommatrix}
&&\hk_N^\mu = \left(\begin{array}{cc} m_N & 0
\\ 0 & m_N
\end{array}\right)\nonumber \\
&& k_4^\mu = \left(\begin{array}{cc} m_N &
{\braket{12}\sbraket{13}\over \braket{23}}
\\ -  {m^2_T\braket{23} \over \braket{12}\sbraket{13}^3} & 0
\end{array}\right)\nonumber \\
&& \hk_1^\mu = \left(\begin{array}{cc} 0 &
-{\braket{12}\sbraket{13}\over \braket{23}}
\\  {m^2_T\braket{23} \over \braket{12}\sbraket{13}} &m_N
\end{array}\right)
\end{eqnarray}
where we have used equations (\ref{Biden4}), (\ref{Biden5}) and
(\ref{momdef}, \ref{momdec34}) to determine the diagonal
components of $\hk_1$ and $k_4$. We also used the expressions in
(\ref{hehk14}) to calculate the off-diagonal components. It is
obvious by simple inspection of (\ref{kmatrix}) that
$\hk_1^{(+)}=0$ and $k_4^{(-)}=0$ with respect to the light-cone
frame defined by the vectors $\hk^\flat$ and $n$. So the
decomposition of the external momenta in terms of $n$ being one of
the two light-cone vectors results in the natural basis
(\ref{kNbasis}) which in turn dictates that one of the two
external momenta i.e. $\hk_1$ has vanishing $\hk_1^{(+)}$
component. This serves in the simplicity of the cubic vertex
(\ref{cubcoupCPpm}) as it is explained in section \ref{seccubic}
and in appendix C.

\renewcommand{\thesection}{C}
\setcounter{equation}{0}
\renewcommand{\theequation}{C.\arabic{equation}}
\section*{Appendix C: Light-cone String Field Theory}\label{C}

In order to derive the form of the vertex in (\ref{cubcoupCPpm})
one has to rely to SFT in the light-cone gauge. The main formulas
for the cubic vertex are given in chapter 11 of
\cite{Green:1987sp}
\begin{equation}\label{lcSFT}
\ket{V_3}= exp \left( -\tau_0 \sum_r {1\over \a_r} + \Delta_B
\right) \ket{0} \delta \left( \sum_r p_r \right)
\end{equation}
where
\begin{equation}\label{DB}
\Delta_B={1\over 2} \sum_{r,s} \sum_{m,n=1}^\infty
\bar{N}^{rs}_{mn} \  \a_{-m}^r \cdot \a_{-n}^s + \sqrt{2\apr}
\sum_{r} \sum_{m=1}^\infty\bar{N}^{r}_{m} \a_{-m}^r \cdot P + \apr
{\tau_0 \over \hat{\a}} P^2
\end{equation}
and $r=1,2,3$ run over the three Hilbert spaces for the particles
of interest\footnote{Notice that we use different metric than
in\cite{Green:1987sp} which means $p^2=m^2$ and moreover they use
$\apr={1\over 2}$}. The definitions of the Neumann matrices are
\begin{eqnarray}\label{Neu}
&& \bar{N}^{r}_{m}= N^{r}_{m}e^{m \tau_0/\a_r} \ , \quad
N^{r}_{m}=
{1\over \a_r} f_m ( -{\a_{r+1}\over \a_r}) \nonumber \\
&& N^{rs}_{mn}= -{m n \a_1 \a_2 \a_3 \over n \a_r + m
\a_s}N^{r}_{m}N^{s}_{n}\nonumber \\
&& f_m(\gamma)= {(-1)^{m-1}\over m!} {\Gamma(m-m \gamma) \over
\Gamma(1-m \gamma)}
\end{eqnarray}
The parameters of the vertex are defined as
\begin{equation}\label{SFTparam}
P= \a_1 k_2 -\a_2 k_1 \ , \quad \tau_0= \sum_{r=1}^3 \a_r \log
|\a_r| \ , \quad \hat{\a}= \a_1 \a_2 \a_3
\end{equation}
The string "lengths" are given by the relation $\a_r= 2 k^{(+)}$
and they satisfy due to momentum conservation the relations
\begin{equation}\label{SFTrel}
\sum_{r=1}^3 {k^2_r \over \a_r}= - {P^2\over \hat{\a}}\ , \quad
\sum_r \a_r=0
\end{equation}
Moreover the above imply that $P$ is cyclic in the Hilbert indices
something which is not obvious in (\ref{Neu}) and that for the
$\a_r$ variables holds a cyclic property with the identification
$\a_4=\a_1$.

With real momenta it is not a'priori possible to set any of the
parameters $\a_r=0$. This is obvious since this would contradict
the on shell condition $2k^{(+)} k^{(+)}-(k^i)^2= m_N^2$. But with
complex momenta we can see that such a choice is possible. We saw
in appendix B that the massive momentum decomposition for the BCFW
setup leads naturally to a light-cone frame where the momentum of
one of the tachyons has $\hk^{(+)}_1=0$ \footnote{A note of
caution. If one of the light-cone momenta of the external states
goes to zero then naively the length of one of the interacting
strings will become zero as well since $0\leq \sigma_r \leq \pi
|\a_r|$. So it is more appropriate to think the case $\a_r=2
k^{(+)}=0$ as an analytic continuation from the case with $a_r\neq
0$ otherwise we would have to discuss a configuration with a
possible singular world-sheet.}.

For our purposes we can put the two tachyons in Hilbert spaces
$r=1,2$ and the reggeon state in $r=3$. We can choose $\a_1=0$.
This in terms of our setup in BCFW means that we can choose the
string light cone frame as in (\ref{kNbasis}). Since only the
state in the third Hilbert space is a reggeon state and the other
two are vacuum states (tachyons), we have to consider only the
behavior of the Neumann coefficient of the third Hilbert space for
this configuration \footnote{ Actually the Neumann coefficients
for the 1st Hilbert space are diverging in general so it is not
obvious if one can extend this reasoning for 3-point functions
which involve other reggeon states.}. We can easily see that the
following identities hold
\begin{eqnarray}\label{SFTiden}
&&\a_2=-\a_3 \ , \quad P= -\a_2 k_1 \, \quad \tau_0 \to 0 \nonumber \\
&& -\tau_0 \sum_r {1\over \a_r}+ \apr{\tau_0 \over \hat{\a}}
P^2=\log|\a_1| \left( 1+\apr k_1^2\right)=0 \nonumber \\
&&N^3_m= {1\over a_3} {(-1)^{m-1}\over m} \ , \quad N^{3}_{mn}=0 \
, \quad \bar{N}^3_m= N^3_m
\end{eqnarray}
where we have used the mass-shell condition for the tachyon
$k_1^2=-{1\over \apr}$. So we see that the Neumann coefficients
which would lead into traces are eliminated in this case. We are
left only with
\begin{equation}\label{DBspecial}
\Delta_B=\sqrt{2\apr} \sum_{m=1}^\infty\bar{N}^{3}_{m} \a_{-m}^3
\cdot P
\end{equation}
Acting on a state as in (\ref{lcstate}) and two tachyons and
restricting to 4-dimensions we indeed get the result in
(\ref{cubcoupCPpm})up to a phase. To prove this we have to use the
oscillator algebra (\ref{oscicom}) and make the substitution
$e^{\pm}\cdot k_1= {1\over 2}e^{\pm}\cdot (k_1-k_2)$ due to
momentum conservation and transversality of the polarization to
$k_3=k_1+k_2$ . The $(-1)^{m-1}$ factors give $(-1)^{N} \times
(-1)^H$. The phase $(-1)^H$ is equivalent to our $O_{abc}^H$.

We could call the above a proof if it was not for the possibly
singular choice of $a_r$ we have made. A more precise derivation
should follow the result of \cite{Ademollo:1974kz}. Specifically
their equation (5.14) includes traces of the reggeon  state which
are given in the string light-cone by $\sum_{i=1}^{24} e^i \cdot
e^i=24$. Definitely the vertex derived in this work differs from
ours. It has as a leading derivative {\it proportional} to the
vertex (\ref{cubcoupCPpm}) but also lower derivative ones since it
corresponds to generic configurations with $k^{(+)}_1 \neq 0$.
Nevertheless the statement we wish to make is that the final
result of the BCFW construction will be the same. In other words
it will agree with the BCFW result using the light-cone vertex in
the particular light-cone frame we have used in this paper. To
prove this, one could choose a different light-cone frame and
repeat the analysis from scratch. This means a different set of
light-cone vectors in place of $\hk_N\ , n$ but also different set
of polarization vectors $\he^{\pm}$. This point definitely
deserves further study.

\renewcommand{\thesection}{D}
\setcounter{equation}{0}
\renewcommand{\theequation}{D.\arabic{equation}}
\section*{Appendix D: A few examples on level contributions}\label{D}

Consider the contribution to (\ref{Hcontr}) of the level 1 state.
There is only one oscillator in this case and the light cone state
is written as
\begin{equation}\label{level1}
\ket{1, k}=  a^{\pm}_{-1}
\end{equation}
This automatically leads to the contribution to the residue of
BCFW from (\ref{Nncontr}) for $N_n=1$
\begin{equation}
({\cal A}_L \times {\cal A}_R)_{1}= {\a(s)+1 \over 1!}
\end{equation}
For this level there is only one contribution and this agrees with
the (\ref{poch}) for $N=1$.

Level $N=2$ is the first level where the particular normalization
of our state will become important. The light-cone states and
their contributions are
\begin{eqnarray}\label{level2}
&& {a^{\pm}_{-2}\over 2 \ 1!} \rightarrow {\a(s)+1 \over 2}
\nonumber \\
&& {(a^{\pm}_{-1})^2 \over 2!} \rightarrow {(\a(s)+1)^2 \over 2}
\end{eqnarray}
Summing all of the above contributions we can easily derive
(\ref{poch}) for $N=2$
\begin{equation}
({\cal A}_L \times {\cal A}_R)_{2}= {(\a(s)+1)(\a(s)+2) \over 2!}
\end{equation}

The next level is $N=3$. There are three light-cone states and we
give their contribution from (\ref{Nncontr})
\begin{eqnarray}\label{level3}
&& {a^{\pm}_{-3}\over 3 \ 1!} \rightarrow {\a(s)+1 \over 3}
\nonumber \\
&& {a^{\pm}_{-2}\ a^{\pm}_{-1} \over 2 \ 1! \ 1!} \rightarrow
{(\a(s)+1)^2 \over 2}
\nonumber \\
&& {(a^{\pm}_{-1})^3 \over 3!} \rightarrow {(\a(s)+1)^3 \over 6}
\end{eqnarray}
Summing all of the above contributions we can easily derive
(\ref{poch}) for $N=3$
\begin{equation}
({\cal A}_L \times {\cal A}_R)_{3}= {(\a(s)+1)(\a(s)+2)(\a(s)+3)
\over 3!}
\end{equation}
 Level 4 goes along the same lines.

    \end{document}